\documentclass[a4paper]{talksvz}
\usepackage{amssymb}
\usepackage{amsthm}
\usepackage{latexsym}
\usepackage{epsfig}
\usepackage{graphicx}
\newcommand{\be}{\begin{equation}}
\newcommand{\ee}{\end{equation}}
\newcommand{\bea}{\begin{eqnarray}}
\newcommand{\nn}{\nonumber}
\newcommand{\eea}{\end{eqnarray}}
\newcommand{\bdm}{\begin{displaymath}}
\newcommand{\edm}{\end{displaymath}}

\begin{document}
\title{\textbf{Geometrical Tachyon Inflation in the Presence of a Bulk Tachyon Field
}\footnote{Talk presented at the "Brane-world Gravity: Progress
and Problems" meeting and workshop, 18th - 29th September 2006,
ICG, University of Portsmouth, UK and at the 12th Conference in
the series "Recent Developments in Gravity" (NEB XII), 29th June -
2nd July 2006, Nafplio, Greece.}}
\author{\textbf{Vassilios Zamarias}}
\date{}
\address{Department of Physics, National Technical University of
Athens,\\
Zografou Campus GR 157 73, Athens,Greece}
\ead{\textbf{zamarias@central.ntua.gr}}
 \vspace{1cm}

\begin{abstract}
 We present the study of the dynamics of the geometrical tachyon field on an
unstable D3-brane in the background of a bulk tachyon field of a
D3-brane solution of Type-0 string theory. We find that the
geometrical tachyon potential is modified by a function of the
bulk tachyon and inflation occurs at weak string coupling, where
the bulk tachyon condenses, near the top of the geometrical
tachyon potential. We also find a late accelerating phase when the
bulk tachyon asymptotes to zero and the geometrical tachyon field
reaches the minimum of the potential.
 \end{abstract}

\section{Introduction}

In open string theory the presence of a tachyon field indicates an
instability in its world-volume. There is strong evidence of a
relation between the full dynamical evolution in string theory and
renormalization group flows on the world-sheet, having both many
features in common. The most profound one is that a world-sheet RG
flow, away from an unstable string background, ends at an infrared
conformal field theory that may generically be expected to be
stable. Similarly, the dynamical process of tachyon condensation
is generically expected to decay into a stable solution of string
theory \footnote{For reviews on open tachyon condensation see
\cite{Taylor:2002uv}.}.

The time evolution of a decaying D-brane in an open string theory
can be described by an exact solution called rolling tachyon
\cite{Sen:2002nu} or S-brane \cite{Gutperle:2002ai}. The
homogeneous decay can be described by perturbing the D-brane
boundary conformal field theory. During this decay described by an
instability of the RG flow on the world-sheet of the string, the
spacetime energy decreases along the RG flow. The end point of
this evolution is to dump the energy released by the tachyon
condensation into the surrounding space, presumably in the form of
closed string radiation, and then relax to its ground state.

Closed string theories are theories of gravity where spacetime is
dynamical. An instability in the spacetime theory implies also an
instability of RG flow, on the world-sheet of the string.
Therefore we expect similar behaviour as in the open string case.
There are however some important differences. The condensation of
a closed tachyon field modifies the asymptotics of spacetime (for
a review on closed tachyon condensation see
\cite{Headrick:2004hz}). Nevertheless it was found that the
spacetime energy decreases along bulk world-sheet RG flows, at
least for the flows for which this statement may be sensibly
formulated. Also, for the case of the closed tachyons,
conservation of energy severely complicates the issue of whether
condensation leads to the true vacuum of the theory, if it has
one.

The Dirac-Born-Infeld action as an effective theory successfully
describes the physics in the world-volume of string theory. In the
case of open string theory, the DBI action was employed to
describe the dynamics of an open tachyon field and in particular
the time evolution of an unstable D-brane by the rolling of a
tachyon field down its potential. When a closed tachyon field is
present in the bulk, the world-volume dynamics is described by a
modified DBI action. In the case of Type-0 string theory, the
world-volume couplings of the tachyon with itself and with
massless fields on a D-brane were calculated
\cite{Garousi:1998fg,Garousi:1999fu}. It was found that the bulk
tachyon appears as an overall coupling function in the DBI action
\cite{Garousi:1998fg,Klebanov:1998yy}.

The Type-0 string theory is an example of a closed string theory,
where the tachyon condensation stabilizes the theory. We are far
from understanding its full dynamics but nevertheless it gives us
some information about the gravitational dynamics of the bulk. It
would be interesting to see what is the effect of the closed
tachyon condensation on the boundary theory.

In this talk we will consider a probe D3-brane moving in the
background of a Type-0 string. We will study a particular D3-brane
bulk solution of this string theory, for which we know an exact
solution at least in the weak string limit. At that limit on the
other hand, open strings do not feel the strong gravity effects
and therefore open string tachyon condensation can take place on
this fixed D3-brane background geometry. However, the probe
D3-brane will be affected by the bulk geometry through a modified
DBI action describing its dynamics, and through the Wess-Zumino
term which encodes the structure of the bulk. Further we will
study the influence if a bulk tachyon field to the cosmological
evolution of the geometrical tachyon.

\section{A Probe D3-Brane Moving in the Background of D3-Branes with a Bulk Tachyon Field}

We consider a general non-conformal string background, with the
presence of a dilaton field, of a stack of D3-branes which are
RR-charged. In particular, we are interested in the movement of a
probe D3-brane in a specific type of the above background, namely
the Type-0  string background
\cite{Polyakov:1998ju,Klebanov1:1998yy,Klebanov:1998yy} in which,
except from the RR fluxes there is also a bulk tachyon field,
coupled to them.

The action of the Type-0 string is given by
\cite{Klebanov1:1998yy}
\begin{eqnarray}\label{action}
S_{10}&=&~\int~d^{10}x \,\, \sqrt{-g}\Big{[} e^{-2\Phi} \Big{(}
 R+4(\partial_{\mu}\Phi)^{2} -\frac{1}{4}(\partial_{\mu}T_{bulk})^{2}
-\frac{1}{4}m^{2}T^{2}_{bulk}-\frac{1}{12}H_{\mu\nu\rho}H^{\mu\nu\rho}\Big{)}
\nonumber \\&& - \frac{1}{4}f(T_{bulk})|F_{5}|^{2} \Big{]}~,
\end{eqnarray} where $F_{5}=dC_{4}$ is the 5-form field strength
of the RR field. The tachyon is coupled to the RR field via the
function $f(T_{bulk})=1+T_{bulk}+\frac{1}{2} T^{2}_{bulk}~$. The
bulk tachyon field appearing in (\ref{action}) is a closed tachyon
field which is the result of GSO projection and there is no
spacetime supersymmetry in the theory. The tachyon field appears
in (\ref{action}) via its kinetic term, its potential and via the
tachyon function $f(T_{bulk})$. The potential term is giving a
negative mass squared term which is signaling an instability in
the bulk. However, it was shown in
\cite{Klebanov:1998yy,Klebanov1:1998yy,Klebanov:1999um} that
because of the coupling of the tachyon field to the RR flux, the
negative mass squared term can be shifted to positive values if
the function $f(T_{bulk})$ has an extremum, i.e.
$f^{\prime}(T_{bulk})=0$. This happens in the background where the
tachyon field acquires vacuum expectation value $T_{bulk,~vac}=-1$
\cite{Klebanov:1998yy,Klebanov1:1998yy}. In this background the
dilaton equation is $\nabla^{2}\Phi=-\frac{1}{4\alpha^{\prime}}
e^{\frac{1}{2}\Phi}T^{2}_{bulk,~vac}~$. This equation is giving a
running of the dilaton field which means that the conformal
invariance is lost, and $AdS_{5}\times S^{5}$ is not a solution.
Therefore, the closed tachyon condensation is responsible for
breaking the 4-D conformal invariance of the theory.  However, the
conformal invariance is restored in two conformal points,
corresponding to IR and UV fixed points, when the tachyon field
gets a constant value. The flow from IR to UV as exact solutions
of the equations of motion derived from action (\ref{action}) is
not known, only approximate solutions exist
\cite{Klebanov:1998yy,Minahan:1998tm,Grena:2000xw}. In these
solutions, the closed tachyon field starts in the UV from
$T_{bulk}=-1$, grows to larger values and passing from an
oscillating phase it reaches $T_{bulk}=0$ at the IR. However, if
the dilaton and tachyon fields are constant, an exact solution of
a D3-brane can be found \cite{Klebanov1:1998yy,Klebanov:1998yy}
 \begin{equation}
ds^{2}_{10}=\frac{1}{\sqrt{H}}\Big{(}-dt^{2}+(d \vec{x})
^{2}\Big{)} +\sqrt{H}\,dr^{2}+ \sqrt{H}\, r^{2}d \Omega ^{2}_{5}~,
\label{bhmetrictyp0}
\end{equation}
where $ H=1+\Big{(} \frac{e^{\Phi_{0}}Q}{2r^{4}}\Big{)}$, Q is the
electric RR charge and $ \Phi_{0}$ denotes a constant value of the
dilaton field. If we define
$L=\Big{(}e^{\Phi_{0}}Q/2\Big{)}^{1/4}$, $H$ can be rewritten as $
H=1+\Big{(} \frac{L}{r}\Big{)}^{4}$.The RR field takes the form
$C_{0123}=\sqrt{1+\Big{(} \frac{r_{0}}{L}\Big{)} ^{4}}\,
\frac{1-H}{H}~$.

The dynamics on the probe D3-brane will be described by the
 Dirac-Born-Infeld plus the Wess-Zumino action,
\be\label{B.I. action}
  S_{probe}=S_{DBI}+S_{WZ}=\tau~\int~d^{4}\xi \,\,
  e^{-\Phi}\sqrt{-det(\hat{G}_{\alpha\beta}+(2\pi\alpha')F_{\alpha\beta}-
  B_{\alpha\beta})}
  +\tau~\int~d^{4}\xi \,\, \hat{C_{4}}~.
\ee{eqnarray}
 The induced metric on the brane is $\hat{G}_{\alpha\beta}=G_{\mu\nu}\frac{\partial x^{\mu}\partial x^{\nu}}
  {\partial\xi^{\alpha}\partial\xi^{\beta}}~$, with similar expressions for $F_{\alpha\beta}$ and
 $B_{\alpha\beta}$.
For an observer on the brane the Dirac-Born-Infeld action is the
volume of the brane trajectory modified by the presence of the
anti-symmetric two-form $ B_{\alpha\beta}$, and world-volume
anti-symmetric gauge fields $ F_{\alpha\beta}$.

In this background, assuming that there are no anti-symmetric
two-form $ B_{\alpha\beta}$ fields and world-volume anti-symmetric
gauge fields $ F_{\alpha\beta}$, in the static
 gauge, $x^{\alpha}=\xi^{\alpha},\alpha=0,1,2,3 $
 using the induced metric we can calculate the bosonic part of the
 brane Lagrangian which reads
\begin{equation}\label{brane Lagr}
\mathcal{L}=\sqrt{A(r)-B(r)\dot{r}^{2}-D(r)h_{ij}\dot{\varphi}^{i}\dot{\varphi}^{j}}
-C(r)~,
\end{equation}
where $h_{ij}d \varphi ^{i} d \varphi^{j}$ is the line
 element of the unit five-sphere, and
\begin{equation}\label{met.fun}
  A(r)=e^{-\Phi_{0}}k(T_{bulk})^{2}H^{-2}(r)~,
  B(r)=e^{-\Phi_{0}}k(T_{bulk})^{2}H^{-1}(r)~,
  D(r)=e^{-\Phi_{0}}k(T_{bulk})^{2}H^{-1}(r)r^2~,
\end{equation}
where $C(r)$ is the $RR$ background. The problem is effectively
one-dimensional and can be solved easily. Since (\ref{brane Lagr})
is not explicitly time dependent and the $\phi$-dependence is
confined to the kinetic term for $\dot{\phi}$, for an observer in
the bulk, the brane moves in a geodesic parameterized by a
conserved energy $E$ and a conserved angular momentum $l^{2}$
which can easily be calculated and together with the expressions
of the momenta give the following form of the DBI action
 \begin{equation}
S_{probe}=\tau\int d^4x \,\,
k(T_{bulk})H^{-1}\Big{[}\sqrt{1-\Big{(}
1-\frac{\alpha^{2}}{(C(r)+E)^{2}}H^{-2}
 \Big{)}}-f^{-1}(T_{bulk})k^{-1}(T_{bulk})\Big{]}\label{dbi2}~,
  \end{equation}
where $\alpha=e^{-\Phi_{0}}k(T_{bulk})$ and $C(r)=\tau
f^{-1}(T_{bulk})H^{-1}+Q_{1}$ with $Q_{1}$ an integration constant
\cite{Papantonopoulos:2000yz} which will be absorbed  in the
energy $E$ while the factor $e^{-\Phi_{0}}$ has been absorbed in
the coupling $\tau$. The function $k(T_{bulk})$ appears because of
the presence of the tachyon field in the bulk. Its form it is
found to be
$k(T_{bulk})=1+\frac{T_{bulk}}{4}+\frac{3T_{bulk}^{2}}{32}+....$
\cite{Klebanov:1998yy,Garousi:1998fg,Garousi:1999fu}. Note that
there is no world-volume coupling of the bulk tachyon field to the
RR fields \cite{Garousi:1999fu}.

The motion of the probe brane in a geodesic with a conserved
energy $E$ and without angular momentum $l^{2}$
(\cite{zamtachyon:2006} for the case including a conserved angular
momentum), can be parameterized by a single scalar field $T_{geo}$
if
 we define
 \begin{equation}
 T_{geo}=\int dH \Big{(}-\frac{L}{4}H^{1/2}(H-1)^{-5/4}\Big{)}
 \label{fieldredf}~.
 \end{equation}
 This relation gives an explicit relation between the radial
 coordinate $r$ and the field $T_{geo}$. Therefore, the scalar
 field $T_{geo}$ has a clear geometrical interpretation in term of
 the coordinate $r$, the distance of the probe brane from the
 D3-branes in the bulk \cite{Kutasov:2003er}.
  If we use the field redefinition
 (\ref{fieldredf}) the action (\ref{dbi2}) becomes
 \begin{equation} \label{TachyonAction}
 S_{probe}= \int d^{4}x \,\, V(T_{geo}) \Big{[}\sqrt{1+g_{\alpha \beta}\partial^{\alpha}T_{geo}
 \partial^{\beta}T_{geo}} -f^{-1}(T_{bulk})k^{-1}(T_{bulk}) \Big{]}\label{dbi3}~,
 \end{equation}
where the only nontrivial component of the metric $g_{\alpha
\beta}$ is the time component and the open tachyon potential is
given by
 \begin{equation}
V(T_{geo})=\frac{\tau k(T_{bulk})}{H(T_{geo})}~.\label{potent}
\end{equation}
The term $\int d^{4}x \,\,
V(T_{geo})f^{-1}(T_{bulk})k^{-1}(T_{bulk})$ in
(\ref{TachyonAction}), is the familiar Wess-Zumino term which is a
function of the geometrical tachyon field and the projection of
the RR field of the bulk on the brane \cite{Billo:1999tv}.

The form of the tachyonic action (\ref{TachyonAction}) implies the
well discussed physical picture, that the movement of the probe
D3-brane in the field of other D3-branes, corresponds to an open
tachyon field rolling down its potential
\cite{Sen:2002nu,Sen:2002an}. The novel feature here is the
presence of the bulk tachyon field $T_{bulk}$. The bulk tachyon
field appears in the potential (\ref{potent}) and if $l\neq 0$ it
also appears in the definition of $T_{geo}$ in (\ref{fieldredf}).
Further we will study what is its effect on the dynamics of the
open tachyon field as it rolls down its potential. For simplicity
we will consider the probe D3-brane moving radially in this
non-conformal string background. From (\ref{fieldredf}) we can
easily derive the following equation \be \frac{dT_{geo}}{dr} =
\sqrt{H(r)} = \sqrt{1+\frac{L^4}{r^4}} \label{diffeq1}~, \ee which
relates the tachyon field $T_{geo}$ to the distance $r$ of the
probe brane from the bulk D3-branes and its potential (for $l=0$)
is given by (\ref{potent}). Because of (\ref{diffeq1}), the
relation between the radial mode and $T_{geo}$ is polynomial,
which means that the explicit form of $T_{geo}(r)$ can not be
found. Indeed, $T_{geo}(r)$ is a combination of elliptic integrals
of the first and the second kind. Nevertheless asymptotically the
differential equation (\ref{diffeq1}) can be evaluated.
\begin{equation}\label{roots2}
\begin{array}{cc}
\begin{array}{ll}
\mbox{IR limit} \\
(r \rightarrow 0)
\end{array}
\left \{ \begin{array}{ll} T_{geo}(r) \sim - \frac{L^2}{r}
\rightarrow - \infty,
\\ \\ \frac{1}{\tau}V(T_{geo}) \sim k(T_{bulk})\frac{L^4}{T_{geo}^4}
\rightarrow 0~,
\end{array}\right.
\begin{array}{ll}
\mbox{UV limit} \\
(r \rightarrow \infty)
\end{array}
\left \{ \begin{array}{ll} T_{geo}(r) \sim r \rightarrow \infty,
\\ \\ \frac{1}{\tau}V(T_{geo}) \sim
k(T_{bulk})\simeq const.
\end{array}\right.
\end{array}
\end{equation}

The presence of the bulk tachyon field through the function
$k(T_{bulk})$ in (\ref{potent}) influences the shape of the
geometrical tachyon potential. According to Sen's conjecture, the
height of the geometrical tachyon potential in its maximum value,
is equal to the tension of the D3-brane. In the UV fixed point
where the bulk tachyon field condenses, $k(T_{bulk})=3/4$,
indicating that the presence of the bulk tachyon is lowering the
D3-brane tension. The bulk tachyon field changes from -1 in the UV
to 0 in the IR, therefore the geometrical tachyon
potential~(\ref{potent}) alters its shape as the geometrical
tachyon rolls down to its minimum. As we will see in the next
section this has important consequences in the cosmological
evolution of the brane-universe.

Observe that in the D3-brane bulk solution because $k(T_{bulk})=1$
in the infrared, the condensation of the open tachyon field
exactly cancels the probe D3-brane tension and then we do not have
any perturbative open string states in the spectrum. We note also
here that if we do not restrict ourselves to the D3-brane exact
solution (\ref{bhmetrictyp0}) and we consider the full system of
equations resulting from the action (\ref{action}), there are
approximate solutions with non-constant tachyon and dilaton fields
\cite{Klebanov:1998yy,Minahan:1998tm}. In these solutions, the
dilaton field in the infrared gets large values and as a
consequence, the string effective coupling  becomes large, making
the whole behaviour of the tachyon condensation in the infrared
questionable.

\section{Coupling to Gravity-Tachyon Cosmology}

The study of the dynamics of the rolling tachyon describing the
time evolution of a decaying D-brane in open string theory,
initiated the development of the tachyon cosmology
\cite{tachyon-cosmology}.

In this section we will study the influence of a bulk tachyon
field to the cosmological evolution of the geometrical tachyon.
Phrasing it in an other way, we will consider the cosmological
evolution of a probe D3-brane as it moves in the gravitational
field of other D3-branes of Type-0. On the probe D3-brane we will
introduce a four-dimensional scalar curvature term. The motivation
for introducing a local gravitational field on the brane is
twofold: strong effects are arising on the brane because of
tachyon condensation, so it is natural to expect the presence of a
gravitational field, and any matter source on the brane, because
of the gravitational field of the bulk, will generate kinetic
gravitational terms on the brane~\cite{Dvali:2000hr}.

To capture the dynamics of the induced gravitational field on the
brane, assuming that it is minimally coupled, we consider the
  DBI action of the geometrical tachyon field coupled to gravity
  \be \label{CosmoTachyonAction} S_{cosmo} =
\int{d^{4}x \sqrt{-g} \Big{(} \frac{R}{16 \pi G} - V(T_{geo})
\sqrt{1+g^{\mu \nu}
\partial_{\mu}T_{geo} \partial_{\nu}T_{geo}}~ \Big{)}}~. \ee
In the above action the geometrical tachyon field acts as a local
matter source on the brane. Assuming that the tachyon field is
described by a homogeneous fluid with $T_{\mu \nu} = pg_{\mu \nu}
+ (p+\rho)u_{\mu}u_{\nu}$ and the fact that $T_{\mu \nu}=
\frac{-2}{\sqrt{-g}} \frac{\partial{S_{cosmo}}}{\partial{g^{\mu
\nu}}}$ we can extract the expressions of the energy density and
pressure. Moreover, for the string background we are considering,
from equations (\ref{dbi2}) and (\ref{TachyonAction}) we have that
\be
\sqrt{1-\dot{T}_{geo}^2}=\frac{\alpha(T_{bulk})}{H}\Big{(}E+\frac{\beta(T_{bulk})}{H}\Big{)}^{-1},
\label{brel} \ee thus obtaining the following expressions for the
energy density and the pressure \bea
\frac{\rho_{tch}(H)}{\tau} &=& E + \frac{\beta(T_{bulk})}{H}~,\label{rtch}\\
\frac{p_{tch}(H)}{\tau} &=&
-\frac{\alpha^2(T_{bulk})}{H^2}\Big{(}E+\frac{\beta(T_{bulk})}{H}\Big{)}^{-1}\label{pres}~.
\eea One should keep in mind here, that in our approach the
geometrical tachyon field we are considering, because of the
identification we did in (\ref{fieldredf}) is a mirage tachyon
field on the probe brane, in the sense that it is an expression of
bulk quantities. For this reason, we use relation (\ref{brel})
which is not derived from the action (\ref{CosmoTachyonAction}).
Also observe that as the brane moves towards the bulk D3-branes,
$\rho_{tch}$ and $p_{tch}$ of (\ref{rtch}) and (\ref{pres}) are
changing, being functions of $r$. The presence of the scalar
curvature term in the action (\ref{CosmoTachyonAction}), assuming
a flat Robertson-Walker metric on the brane, leads to the
Friedmann equation $\mathcal{H}^2_{tch}=\frac{8\pi
G}{3}\rho_{tch}~$, which gives a cosmological evolution because of
the presence of the gravitational field on the brane.

As the probe brane is moving in the background string theory, it
will also experience the effect of the bulk gravitational field.
This effect can be calculated with the techniques of mirage
cosmology~\cite{Kehagias:1999vr,Papantonopoulos:2000yz,mirage-cosmology}.
The presence of the Type-0 string background induces on the probe
brane~\cite{Papantonopoulos:2000yz}, a four-dimensional metric
\begin{equation}\label{fin.ind.metric}
d\hat{s}^{2}=-d\eta^{2}+g(r(\eta))(d\vec{x})^{2},
\end{equation}
which is the standard form of a flat expanding universe where
$\eta$ is the cosmic time defined by
$d\eta=\frac{H^{-5/4}\alpha(T_{bulk})}{\Big{(}\beta(T_{bulk})H^{-1}+E\Big{)}}dt~$.
Moreover setting $g=a$ we get~\cite{Papantonopoulos:2000yz}
\begin{equation}\label{dens}  \Big{(}\frac
{\dot{a}}{a}\Big{)}^{2}= \frac{\tau}{L^{2}}\Big{(}
\frac{H-1}{H}\Big{)}^{5/2}\Big{(}\alpha(T_{bulk})^{-2}(\beta(T_{bulk})+EH)^{2}-1
\Big{)}~,
\end{equation}
 where the dot stands for derivative with respect to cosmic time. The
 right hand side of (\ref{dens}) can be interpreted in terms of an
 effective or ``mirage" matter density on the probe brane
 \begin{equation}\label{rmir} \rho_{mir}=
\frac{3\tau}{8\pi G L^{2}}\Big{(}
\frac{H-1}{H}\Big{)}^{5/2}\Big{(}\alpha(T_{bulk})^{-2}H^{2}\,\Big{(}E+\frac{\beta(T_{bulk})}{H}\Big{)}^{2}-1
\Big{)}~.
\end{equation} We can also calculate the ``mirage" pressure using
$\frac{\ddot{a}}{a}= \Big{[}1+\frac{1}{2}a\frac{\partial}
{\partial a}\Big{]}\frac{8\pi G}{3}\rho_{mir}$ and setting  the
above equal to $-\frac{4\pi G }{3}(\rho_{mir}+3p_{mir})$, we can
calculate the ``mirage" pressure $p_mir$. Then, as the brane is
moving in the gravitational field of the bulk, because of this
motion~~\cite{Kehagias:1999vr}, there will be a cosmological
evolution on the brane described by the Friedmann equation
$\mathcal{H}^2_{mir}=\frac{8\pi G}{3}\rho_{mir}~$.

\begin{figure}
\begin{center}
\includegraphics[scale=0.67]{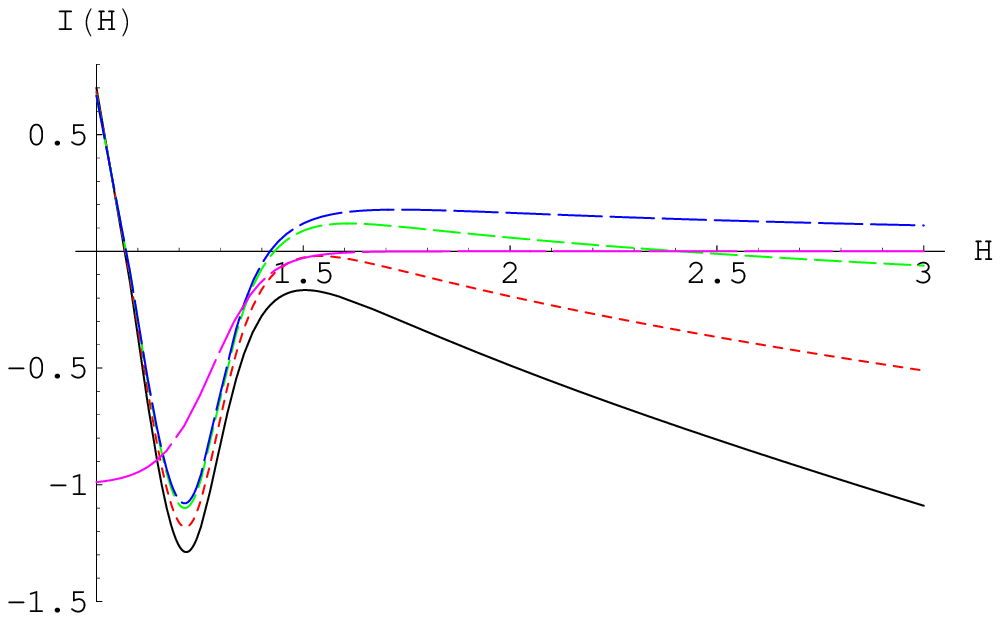}
\hspace{0.05cm}%
\includegraphics[scale=0.85]{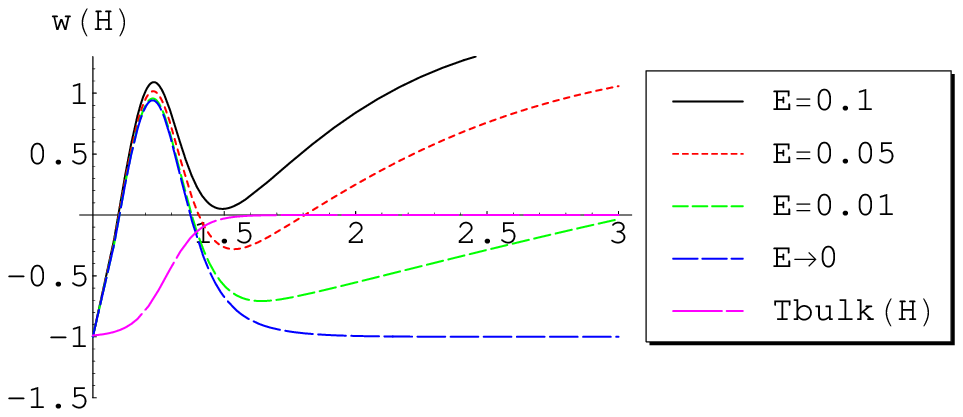}
\caption{The inflation parameter $I(H)$ and the equation of state
parameter $w(H)$ as functions of $H$ for $\zeta=8$, $\sigma=1.28$
and for different values of $E$ in $\tau$ units.}
\label{InflationPlot}
\end{center}
\end{figure}

Therefore, as the geometrical tachyon rolls down its potential it
feels two effects: the gravitational field of its own and the
gravitational field of the bulk D3-branes. Our basic assumption of
the probe approximation allows us to assume, because the D3-brane
as it moves does not backreact with the background, that the above
two contributions give an additive effect on the cosmological
evolution of the probe brane,  and hence it is described by the
Friedmann equation

\be \mathcal{H}^2_{probe}= \frac{8\pi
G}{3}\Big{(}\rho_{tch}+\rho_{mir} \Big{)}\label{fried}~. \ee Also
the Raychaudhury equation $\mathcal{H}_{probe}$ can be calculated.
The inflationary parameter $I(H)$ can be defined as $I(H) =
\mathcal{H}^2_{probe} +\dot{\mathcal{H}}_{probe}^{2}$ and is found
to be \bea I(H)&=& -\frac{4\pi G
\tau}{3}\frac{(H-1)^{5/4}}{H^{5/2}}
\Big{\{}\frac{4\sqrt{3}}{\alpha(T_{bulk})}\,\frac{\beta(T_{bulk})H^{5/4}}{\sqrt{8\pi
G \,L^2}}\,\Big{(}E+\frac{\beta(T_{bulk})}{H}\Big{)}^{1/2}\nn \\
&~&\times\Big{[}1-\frac{\alpha(T_{bulk})^{2}}{H^{2}}\Big{(}E+\frac{\beta(T_{bulk})}{H}\Big{)}^{-2}\Big{]}^{1/2}
\nonumber \\
&-&2
\Big{(}E+\frac{\beta(T_{bulk})}{H}\Big{)}\frac{H^{5/2}}{(H-1)^{5/4}}\Big{[}1-\frac{3}
{2\pi G\,L^2}\,\,\frac{E\,(H-1)^{5/2}}{\alpha(T_{bulk})^{2}\,H^{1/2}}\Big{]} \nonumber \\
&-&\frac{3}{4\pi
G\,L^2}(H-1)^{1/4}\Big{[}\frac{H^{2}}{\alpha(T_{bulk})^{2}}
\Big{(}E+\frac{\beta(T_{bulk})}{H}\Big{)}^{2}-1\Big{]}\Big{(}H-6\Big{)}\Big{\}}~.
\eea The inflationary parameter $I(H)$ depends on the value of
$T_{bulk}$. As we discussed before, we do not know the exact
variation of the bulk tachyon field from UV to IR. The existing
approximate solutions are valid only in the vicinity of the fixed
points and they can not give us a cosmological evolution from
large to small distances. However, we know that the bulk tachyon
field varies from the UV value $T_{bulk}=-1$ to the IR value
$T_{bulk}=0$. We will simulate then this variation with a smooth
function \be
T_{bulk}(H)=\frac{1}{2}\Big{[}\textnormal{Tanh}\Big{(}{\zeta(H-\sigma)}\Big{)}-1\Big{]}~,
\ee where the parameter $\zeta$ controls how steep is the
variation, while $\sigma$ indicates when the transition from -1 to
0 occurs. Using this function, in Fig.~\ref{InflationPlot} we have
plotted the inflationary parameter as a function of $H$ and for
various values of the energy $E$ (in units of $\tau$). The
cosmological evolution of the brane-universe starts with an early
inflationary phase near the value $T_{bulk}=-1$, where the bulk
tachyon field condenses and the string coupling is weak, and as
the bulk tachyon field gets larger values, there is an exit from
inflation and a late acceleration phase as the bulk tachyon field
approaches $T_{bulk}\rightarrow 0$.

We can see the cosmological evolution on the brane-universe using
the geometrical tachyon picture. As we discussed in the Sect. 2,
the geometrical tachyon rolling down its potential has two
different asymptotic behaviours. At the UV it starts with
$T_{geo}=\infty$ at the top of the potential, and rolls down to
$V(T_{geo})=0$ in the IR where $T_{geo}=-\infty$. The transition
to the two regimes occurs where $r=L$ or $H=2$. On the other hand,
the background string solution (\ref{bhmetrictyp0}), is reliable
near the UV and IR fixed points in which $H=1$ and
$H\rightarrow\infty$ respectively. As we can see in
Fig.~\ref{InflationPlot}, there is a choice of parameters for
which the early inflationary phase and the exit from it occurs
around $H=1$ which corresponds to the top of the geometrical
tachyon potential. The late inflationary phase occurs for large
$H$ values, where the bulk tachyon field has decoupled, and this
happens in the bottom of the geometrical tachyon potential.

We can also define the equation of state parameter $w(H)=p_{probe
}/\rho_{probe}$. Then, using (\ref{rtch}), (\ref{pres}),
(\ref{rmir}) and the expression of $p_{mir}$ we can plot $w$ as a
function of $H$ and for various values of the energy. We see in
Fig.~\ref{InflationPlot} that it starts with the value $w=-1$ in
the early inflationary phase, then it gets positive values and
finally relaxes again to $w=-1$ in the late accelerating phase.
This picture is appealing, because the whole cosmological
evolution is driven by the dynamics of the theory, without any
dark matter or dark energy.

If we switch off the gravitational field on the moving brane then,
the mirage effect~\cite{Papantonopoulos:2000yz}  gives only the
late accelerating  phase as the probe brane is approaching the
bulk. If we switch off the mirage contribution then the
brane-universe has the early inflationary phase, where the tachyon
field condenses, near the top of the geometrical tachyon
potential. If we decouple the bulk tachyon field from the start,
having a probe brane moving in the background of other D3-branes
then, we have a very short inflationary period in the top of the
geometrical tachyon potential. This can be attributed to the
presence of the gravitational field on the brane.

A remark concerning the consistency of our approach. Let us recall
what happens in the DGP model~\cite{Dvali:2000hr}. We have a
static brane in a time dependent five-dimensional pure
gravitational bulk. As it was showed in \cite{Kofinas:2001es}, the
introduction of a four-dimensional scalar curvature term on the
brane, simply redefines the energy momentum tensor on the brane.
If we go to the picture in which the brane is moving and the bulk
is static, it can be shown~\cite{csaki}, that the $R$ term has the
same effect, it redefines the energy momentum tensor in the
junctions conditions which play the r$\hat{o}$le of the equations
of motion of the moving brane. In both pictures, there is an
effective four-dimensional Einstein equation which describes
consistently the cosmological evolution on the brane.

In the case of a Dp-brane moving in the background of other
Dp-branes the situation is much more
difficult~\cite{Kiritsis:2003mc}. The Dp-branes of the bulk are
solutions of a complicated usually string theory and the only
information we can get on the brane is only through the DBI
action. For this reason we use the rolling tachyon picture which
is described by a simple DBI action and the basic assumption of
the probe brane approximation. There is no backreaction between
the brane and the bulk. As the brane moves in the gravitational
field of the background branes it does not disturb this
background. This approach led us to the Friedmann equation
(\ref{fried}).

\section{Conclusions and Discussion}

In our talk we provided some evidence that closed tachyon
condensation may have some important consequences on the
cosmological evolution of the boundary theory. Of course we do not
fully understand the dynamics of the closed tachyon condensation
and its connection to the gravitational dynamics, but nevertheless
we provided an example in which the condensation of the bulk
tachyon, except that it stabilizes the bulk theory, it is
responsible for the inflation on the boundary theory.

More work is needed to understand the phenomenological aspects of
the inflation on the brane, like how long it lasts, what are the
scalar perturbations produced during inflation, what is the
mechanism for reheating. Can the condensation of the bulk tachyon
provide the energy needed for the reheating? However, there is a
drawback in these considerations, because the relation
(\ref{brel}) indicates that the kinetic energy of the geometrical
tachyon field is not small and it can not be ignored compared to
unity and hence slow-roll inflation can not be applied. This can
be understood because of the strong bulk effects of the bulk
tachyon condensation.

\ack

This work was done in collaboration with Eleftherios
Papantonopoulos and Ioanna Pappa. Work supported by (EPEAEK
II)-Pythagoras (co-funded by the European Social Fund and National
Resources).

\end{document}